\documentclass[aps,prl,twocolumn,showpacs,groupedaddress]{revtex4}
\usepackage{graphicx}

\newlength\figurewidth
\begin{document}
\newcommand{\rem}[1]{}

\title{  Admittance and noise in an electrically driven nano-structure: Interplay between quantum coherence and statistics }

\author{ Hee Chul Park}\author{Kang-Hun Ahn}
\affiliation{ Department of Physics, Chungnam National University,
 Daejeon 305-764, Republic of Korea}

\date{ \today}

\begin{abstract}
We investigate the interplay between the quantum coherence and
statistics in electrically driven nano-structures. We obtain
expression for the admittance  and the current noise for a driven
nano-capacitor in terms of the Floquet scattering matrix and derive
a non-equilibrium fluctuation-dissipation relation. As an interplay
between the quantum phase coherence and the many-body correlation,
the admittance has peak values whenever the noise power shows a step
as a function of near-by gate voltage.
 Our theory is demonstrated by calculating the admittance and noise of
driven double quantum dots.
\end{abstract}
 \maketitle

 Quantum dynamics in time-dependent potential is an important topic
because of its unrevealed non-equilibrium  phenomena as well as the
increasing demand of manipulating coherent electronic states in
quantum information science\cite{feve,gabelli}. While the
time-dependent potentials in diffusive conductors are known to have
destructive role on the quantum coherence of electrons, it is not
always obvious which role is played on the quantum coherence by the
time-periodic potential in ballistic nano-structures.
  {\it Quantum coherence} and {\it quantum many-particle statistics} of
relevant particles are key concepts to understand the electron
transport properties.  For instance, the current noise, which is
basically two-particle property, contains the information on the
quantum many-particle statistics\cite{blanter}. Furthermore, the
noise of a coherent conductor in the presence of time-periodic
external field is known to be sensitive to the phase of transmission
amplitudes\cite{Levitov,Lamacraft,camalet}.

In this work, we study how the interplay between the quantum
coherence and the statistics affects on the transport properties in
the presence of time-periodic external potential. We particularly
concentrate on the noise of {\it quantum} displacement current
through the driven capacitor, because in usual conductors, Shot
noise (in {\it classical} granular nature of electrons) dominates
other types of noises. By this means, we derive the analytic
expression for the admittance and noise formula in terms of the
Floquet scattering matrix for the driven nano-capacitor.

The linear response of the capacitor is described by the admittance
$g(\omega)=I(\omega)/V(\omega)$ which relates the displacement
current $I(\omega)$ to the applied voltage between capacitor
$V(\omega)$. The expression which relates equilibrium admittance and
the noise power $S(\omega)$ to the scattering matrix has been
obtained by B\"uttiker and his coworkers \cite{buttiker,nigg}. Here
we generalize the expression to the case of the non-equilibrium
states generated by time-periodic potential of frequency $\Omega \gg
\omega$. It will be shown here that, as in the case of equilibrium
capacitors, the admittance of the driven capacitor can be also
understood in terms of the time delay of electrons near Fermi level.
Meanwhile, the non-equilibrium current noise power can not be
understood within single-particle picture. The noise power shows
step structure as a function of the nearby gate voltage, which is
associated with the opening of temporal channels in the lead.

Let us begin by introducing a model for the system of a biased
dynamic capacitor.(See Fig. 1.)
 A time-dependent potential with frequency $\Omega$ is applied in a
nano-structure (box of dotted line) which is connected to a
mesoscopic conductor.
 The chemical potential of the mesoscopic
conductor is controlled by a nearby gate voltage $V_{g}$. The
oscillating nano-structure and the mesoscopic conductor are in a
loop enclosing a time-periodic magnetic flux $\Phi(t)$. The bias
voltage $V(t)$ between nano-structure and external lead is induced
as an electromotive force along the loop, $V(t)=d\Phi/dt=
V(\omega)e^{i\omega t}$+c.c.

\begin{figure}
\includegraphics[width=0.8\figurewidth]{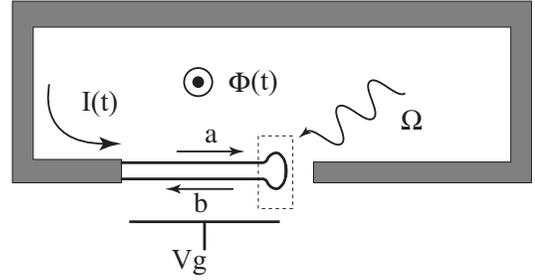}
 \caption{
A time-periodic driven nanostructure (dotted box) is connected to a
mesoscopic conductor. For bias voltage, an external loop driven by
slow-varying flux $\Phi(t)=\Phi_{\omega}\cos\omega t$ is connected.
}

   \label{system-admit}
\end{figure}

We assume the mesoscopic conductor is spatially one-dimensional and
the electrons are non-interacting spin-polarized gas for simple
presentation. Generalization of our results to the case of
multi-channel conductors with spin is straightforward. The
Hamiltonian for the conductor reads
\begin{eqnarray}
H_{0}=\int d\epsilon ~~ \epsilon ~~ (
a^{\dagger}(\epsilon)a(\epsilon)+b^{\dagger}(\epsilon)b(\epsilon) ),
\end{eqnarray}
where $a(\epsilon) (b(\epsilon))$ is the annihilation operator for
the incoming (outgoing) electron to the driven nano-structure;
$\{a^{\dagger}(\epsilon),a(\epsilon^{\prime})\}=
\{b^{\dagger}(\epsilon),b(\epsilon^{\prime})\}=
\delta(\epsilon-\epsilon^{\prime})$,
$\{a(\epsilon),a(\epsilon^{\prime})\}=
\{b(\epsilon),b(\epsilon^{\prime})\}=
\{a(\epsilon),b^{(\dagger)}(\epsilon^{\prime})\}=0$.

 A time-periodic steady states are formed in the conductor and the
driven nano-structure
\begin{eqnarray}
i\hbar \frac{\partial}{\partial t} |\psi_{\epsilon}(t)\rangle
=(H_{0}+H_{\Omega}(t)+H_{c})|\psi_{\epsilon}(t)\rangle,
\label{schroedinger}
\end{eqnarray}
where $H_{\Omega}$ is the time-periodic Hamiltonian for the electron
in the driven nano-structure and $H_{c}$ denotes the coupling
between the lead and the nano-structure.
$|\psi_{\epsilon}(t)\rangle$ is formed by incoming electron state of
energy $\epsilon$ and the linear combinations of its scattered
states with energy $\epsilon + n \hbar\Omega$. We assume the
external metallic lead is big enough so that it can be considered as
a reservoir in thermal equilibrium. The occupation number for the
incoming electron state of energy $\epsilon$ is given by Fermi-Dirac
distribution function $f(\epsilon)$.

Let us consider the linear response of the electric current to the
time-dependent magnetic flux. The perturbing time-dependent
Hamiltonian is $H_{\omega}(t)=\Phi(t) \hat{I} $ where $\hat{I}$ is
the current operator
\begin{eqnarray}
\hat{I}=\frac{e}{h} \int d\epsilon d\epsilon^{\prime} ~ (
a^{\dagger}(\epsilon)a(\epsilon^{\prime})-b^{\dagger}(\epsilon)b(\epsilon^{\prime})).~
\label{current-operator}
\end{eqnarray}
In fact, the above current formula is an approximation where the
current value is taken as the spatial average value over the length
of the conductor $L$.  This approximation holds when the relevant
frequency scale is much smaller than the Fermi velocity divided by
the length of the conductor $L$, i.e. $\omega,\Omega \ll v_{F}/L$,
so that the relevant wavenumber $k$ satisfies $|k-k_{F}|L\ll 1$, and
$k_{F}L \gg 1$ where the rapidly oscillating phase terms are washed
out.

 The adiabatic turning on of $\Phi(t)=\Phi_{\omega}\cos\omega t
e^{0^{+}t}$ gradually deforms $|\psi_{\epsilon}(t)\rangle$ to
$|\Psi_{\epsilon}(t)\rangle=\int d\epsilon^{\prime}C_{\epsilon
\epsilon^{\prime}}(t)|\psi_{\epsilon^{\prime}}(t)\rangle.$
The coefficients $C_{\epsilon \epsilon^{\prime}}(t)$ are determined
by solving the Schr\"odinger equation for the time-dependent
Hamiltonian $H(t)=H_{0}+H_{\Omega}(t)+H_{c}+H_{\omega}(t)$. By
employing a perturbation expansion of
$C_{\epsilon\epsilon^{\prime}}$ in terms of $\Phi_{\omega}$, we get
the first order term $
C^{(1)}_{\epsilon\epsilon^{\prime}}=\frac{\Phi_{\omega}}{i\hbar}\int^{t}_{-\infty}dt^{\prime}\cos\omega
t^{\prime}
\langle\psi_{\epsilon^{\prime}}(t^{\prime})|\hat{I}|\psi_{\epsilon}(t^{\prime})\rangle$.

 Up to first order of $\Phi_{\omega}$, the $\omega$ component of the displacement current, $I^{(1)}_{\epsilon,\omega}$, is
 obtained through
\begin{eqnarray}
\label{current-formula}
\langle\Psi_{\epsilon}(t)|\hat{I}|\Psi_{\epsilon}(t)\rangle &=&\int
d\epsilon^{\prime}\langle\psi_{\epsilon}(t)|\hat{I}|\psi_{\epsilon^{\prime}}(t)\rangle
C^{(1)}_{\epsilon\epsilon^{\prime}}(t)+{\rm c.c.} \nonumber \\
&\approx& I^{(1)}_{\epsilon,\omega}e^{i\omega t}+
I^{(1)}_{\epsilon,-\omega}e^{-i\omega t} .
\end{eqnarray}

Floquet theorem says, the eigenstates $|\psi_{\epsilon}(t)\rangle$
of time-periodic Hamiltonian  can be written in terms of
time-independent basis $|\phi^{(l)}_{\epsilon}\rangle$ as
$|\psi_{\epsilon}(t)\rangle=\exp(-i\frac{\epsilon}{\hbar}
t)\sum_{l=-\infty}^{\infty}e^{-i l \Omega t}
|\phi^{(l)}_{\epsilon}\rangle \label{Floquet-wavefunc}$.
Using Eqs.(\ref{current-operator},\ref{current-formula}), one can
relate $I^{(1)}_{\epsilon,\omega}$ to the Floquet eigenstates
$|\phi^{(l)}_{\epsilon}\rangle$, which is useful for further
calculations. The thermal-averaged displacement current is given by
$I(\omega)=\int d\epsilon ~f(\epsilon)~ I^{(1)}_{\epsilon,\omega}$
because the incoming electrons are
 from the reservoir in equilibrium.

The admittance is given by the induced displacement current
$I(\omega)$ divided by the applied voltage
$V(\omega)=i\omega\Phi_{\omega}/2$. After some algebra we find;
\begin{eqnarray}
\nonumber g(\omega)&=&\frac{1}{i\omega}\sum_{m}\int d\epsilon
d\epsilon^{\prime}
|\sum_{l}\langle\phi^{(l)}_{\epsilon}|\hat{I}|\phi^{(l+m)}_{\epsilon^{\prime}}\rangle|^{2}
\\ \nonumber &\times&
\frac{f(\epsilon)-f(\epsilon^{\prime})}{\epsilon-\epsilon^{\prime}-\hbar\omega-m\hbar\Omega+i0^{+}}.
\end{eqnarray}
Here, $|\omega|<|\Omega|$ is assumed.

 In a quantum conductor with a
time-periodic scatterer, the scattering relation between the
incoming electron of energy $\epsilon$ and the out-going electron of
energy $\epsilon^{\prime}$ is given by Floquet scattering
matrix\cite{moskalets}, $S_{F}(\epsilon^{\prime},\epsilon)$.  The
Floquet state in the quantum conductor can be written
\begin{eqnarray}
|\phi_{\epsilon}^{(l)}\rangle=(a^{\dagger}(\epsilon)\delta_{l0}+
s_{l}(\epsilon)b^{\dagger}(\epsilon+l\hbar\Omega))|0\rangle,
\end{eqnarray}
where $s_{l}(\epsilon)=S_{F}(\epsilon+l\hbar\Omega,\epsilon)$. The
unitarity of the scattering matrix gives
$\sum_{l}|s_{l}(\epsilon)|^{2}=1$ and its time-reversal symmetry
gives $s_{-l}(\epsilon+l\hbar\Omega)=s_{l}(\epsilon)$. The current
matrix element in the Floquet basis simply reads
\begin{eqnarray}
\langle\phi^{(l)}_{\epsilon}|\hat{I}|\phi^{(l+m)}_{\epsilon^{\prime}}\rangle
=\frac{e}{h}\Big(
\delta_{l,0}\delta_{m,0}-s^{*}_{l}(\epsilon)s_{l+m}(\epsilon^{\prime})
\Big)
\end{eqnarray}

The real part of the admittance $g^{\prime}(\omega)$ is now written
using $\frac{1}{x+i0^{+}}=P\frac{1}{x}-i\pi\delta(x)$;
\begin{eqnarray}
\nonumber g^{\prime}(\omega)&=&\frac{e^{2}}{2h}\sum_{m}\int
d\epsilon
\Big|\delta_{m0}-\sum_{l}s_{l+m}(\epsilon)s_{l}^{*}(\epsilon+\hbar\omega+m\hbar\Omega)\Big|^{2}
\\&\times&
\frac{f(\epsilon)-f(\epsilon+\hbar\omega+m\hbar\Omega)}{\hbar\omega}
\label{gw-scattering}
\end{eqnarray}
The above result is partly confirmed by the fact that if the high
frequency $\Omega$-radiation were not there,  then $s_{l}\propto
\delta_{l,0}$ and Eq.(\ref{gw-scattering}) is equivalent to Eq.(2)
in Ref.\cite{buttiker}. It is worth noting that the admittance is a
quantity governed by the electron near the Fermi level. At low
frequency and zero temperature, the admittance is approximated by
\begin{eqnarray}
g^{\prime}(\omega) \approx
\frac{e^{2}}{2h}\omega^{2}(\tau_{d}^{2}(E_{F})+\tau^{2}_{p}(E_{F})),
\label{gw-approx}
\end{eqnarray}
where $\tau_{d}$ is the phase delay time\cite{wigner} of the
electron in the nano-structure which is defined as
$\tau_{d}(\epsilon)=-i\hbar\sum_{l}s^{*}_{l}(\epsilon)\frac{ds_{l}(\epsilon)}{d\epsilon}$
and $\tau_{p}$ (relatively smaller than $\tau_{d}$ for weaker
electrical driving ) is the non-equilibrium photo-assisted phase
delay time defined by $\tau^{2}_{p}(\epsilon)=\sum_{m\neq 0}|\hbar
\sum_{l}s_{l+m}(\epsilon-m\hbar\Omega)\frac{ds_{l}^{*}(\epsilon)}{d\epsilon})|^{2}$.

 Now we turn to the (non-symmetrized)
current noise $S(\omega)$ defined by {\begin{eqnarray}
S(\omega)\delta(\omega+\omega^{\prime})=\frac{1}{\pi}\int {\rm dt}
{\rm dt}^{\prime}e^{i\omega t}e^{i \omega^{\prime}
t^{\prime}}\langle \delta \hat{I}(t)  \delta \hat{I}(t^{\prime})
 \rangle
\end{eqnarray}
   where $\delta I(t)=\hat{I(t)}-\langle\hat{I}(t)\rangle$ and
$\hat{I}(t)=e^{iH_{0}t/\hbar}\hat{I}e^{-iH_{0}t/\hbar}$. Again
$\omega$,$\omega^{\prime}$ are assumed low enough that terms of
higher harmonics involving $\delta(\omega+\omega^{\prime}+n\Omega)$
($|n|>$1) vanish. Here the average $ \langle \cdots \rangle$ means
the spatial average after both quantum mechanical and statistical
average over many particle states
$\frac{1}{2^{(N-1)/2}}\prod_{\epsilon_{j}}\Big[a^{\dagger}(\epsilon_{j})+
\sum_{l}
s_{l}(\epsilon_{j})b^{\dagger}(\epsilon_{j}+l\hbar\Omega)\Big]|0\rangle
$
 with thermodynamic weighting factor
$e^{-\beta\sum_{j}(\epsilon_{j}-E_{F} )}$.  The current correlation
$\langle\hat{I}(\omega)\hat{I}(\omega')\rangle$ is written in terms
of the incoming and outgoing particle operators,
($a,a^{\dagger},b,b^{\dagger}$).
The calculation can be easily done by projecting the total
many-particle states into incoming particle states. In the projected
basis, $b(\epsilon)$ is replaced with
$\sum_{l}s_{l}(\epsilon)a(\epsilon+l\hbar\Omega)$.

The correlation between the outgoing particles comes from the
exchange correlation among incoming particles.
 After some algebra, the non-symmetrized noise power of the driven conductor is given
by
\begin{eqnarray}
\nonumber S(\omega)&=&\frac{e^{2}}{h}\sum_{m}\int d\epsilon \big|
\delta_{m,0}-\sum_{l}s_{l+m}^{*}(\epsilon)s_{l}(\epsilon+\hbar\omega+m\hbar\Omega)
\big|^{2}\\&\times&
f(\epsilon)(1-f(\epsilon+\hbar\omega+m\hbar\Omega)).
\label{sw-scattering}
\end{eqnarray}
 By keeping only $m=0$ term, we recover the result for the case of
time-independent potential.\cite{buttiker}

Eqs.(\ref{gw-scattering}) and (\ref{sw-scattering}) are the main
results of this work. One can notice that current noise power
$S(\omega)$ is related to the admittance $g^{\prime}(\omega)$ via
\begin{eqnarray}
S(\omega)-S(-\omega) =2\hbar\omega g^{\prime}(\omega)
\end{eqnarray}
Lesovik and Loosen\cite{lesovik} has shown the above
fluctuation-dissipation relation is valid in a non-equilibrium case
where the particle current flow at small finite bias. In this
expression, the admittance $g^{\prime}(\omega=0)$  is essentially
the derivative of $S(\omega)$ at $\omega=0$. Here, we prove the
fluctuation-dissipation relation for the photo-assisted
non-equilibrium case.

The noise power in Eq.(\ref{sw-scattering}) can be divided into two
different parts, $S(\omega)=S_{0}(\omega)+S_{P}(\omega)$.  They are
 the equilibrium noise, $S_{0}(\omega)$ ($m=0$) and non-equilibrium
noise, $S_{P}$, ($m\neq 0$).
 At low frequency and low temperatures, the non-equilibrium noise $S_{P}(\omega)$ is
more important than equilibrium noise $S_{0}(\omega)$. The
equilibrium noise is proportional to $\omega^3$ and the
non-equilibrium noise is proportional to $\omega^2$. So there is
always a frequency regime where the non-equilibrium noise
$S_{P}(\omega)$ dominates the equilibrium noise $S_{0}(\omega)$ at
low frequencies.

To demonstrate our theory, we consider electrically driven double
quantum dots (DQDs) connected to a single (spatial) channel lead. We
employ the Floquet scattering theory based on tight-binding
approximation to obtain the Floquet scattering matrix element
$s_{l}(\epsilon)$\cite{ahn-physe}.

In the tight-binding model, the localized states in the dots and the
leads are created by $d^{\dagger}_{1(2)}$ and $c^{\dagger}_{j}$
(j=-1,-2,-3,...), respectively. The Hamiltonian for the lead and the
dot-lead coupling is given by
 $H_{0}= -\frac{V_{0}}{2}\sum_{j< -1}
( c_{j+1}^{\dagger}c_{j}+c_{j}^{\dagger}c_{j+1}) $ and
$H_{c}=-\gamma\sum_{i=1}^{2}(d_{i}^{\dagger}c_{-1}
+c_{-1}^{\dagger}d_{i})$. $V_0/2$ is the hopping parameter for the
leads which controls the kinetic energy. $\gamma$ is the tunnel
coupling between the dots and the lead. The Hamiltonian for the
driven double dots in the base of localized state is
\begin{eqnarray}
 H_{\Omega}(t)=\frac{1}{2}\left(
   \begin{array}{cc}
     \Delta -eV_{\Omega}\cos\Omega t& -\Delta_{0} \\
     -\Delta_{0} & -\Delta+eV\cos\Omega t \\
   \end{array}
 \right),
\end{eqnarray} where $\Delta$ and $\Delta_{0}$ is the asymmetry
energy and tunnel splitting energy of the double-quantum dot. The
scattering matrix elements $s_{l}(\epsilon)$ are obtained by a phase
matching method using the incoming state,
$a^{\dagger}(\epsilon)|0\rangle
=\frac{1}{\sqrt{v(\epsilon)}}\sum_{j}e^{ik_{0}j}c^{\dagger}_{j}|0\rangle$
and its outgoing states $b^{\dagger}(\epsilon+l\hbar\Omega)|0\rangle
=\frac{1}{\sqrt{v(\epsilon+l\hbar\Omega)}}\sum_{j}e^{-ik_{l}j}c^{\dagger}_{j}|0\rangle$,
where $\epsilon+l\hbar\Omega=-V_{0}\cos k_{l}$ and $v(\epsilon)$ is
the group velocity.

\begin{figure}
\includegraphics[width=0.7\figurewidth]{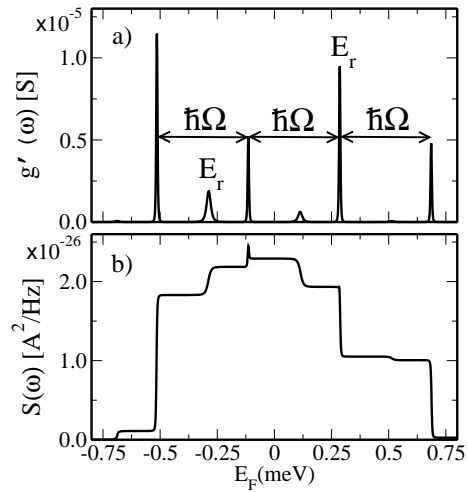}
 \caption{
a) Admittance of the driven double quantum dot in a capacitor as a
function of the Fermi energy at zero temperature.
 b) Non-symmetrized current noise power $S(\omega)$ for the same
 system
as a function of the Fermi energy of the external lead. $E_{F}=0$ is
the case when the Fermi level matches with the center of the two
eigenenergies of the DQD. The parameters in use are
$\gamma$=0.16meV, $\Delta_0$=0.3meV, $\Delta$=0.4meV,
$\Omega$=60.7GHz, $eV_\Omega$=0.3meV, and
$\omega$=4.55GHz,$V_{0}$=5.1meV, respectively.  } \label{ns-noise}
\end{figure}

We show the admittance of driven double quantum dots in Fig. 2 a).
It shows peak structure as a function of Fermi energy. The
admittance has peaks when the Fermi energy matches with the resonant
energy levels of the DQD as well as photon side bands,
$E_{F}\approx E_{r} \pm\hbar \Omega$ where $E_{r}$ is the Floquet
eigenvalue of $H_{\Omega}(t)$. For weak driving, $E_{r}$ is the
energy eigenvalues for DQD $E_{r}\approx \pm
\frac{1}{2}\sqrt{\Delta^{2}+\Delta_{0}^{2}}$.
 {\it Why does it show peaks?}
The nonzero admittance of a capacitor is due to the time delay of
the electrons at the capacitor. Since only states near the Fermi
level are excited by oscillating magnetic flux of low frequency
$\omega$, the admittance is naturally given by the Fermi level
quantity. So, the peak values of the dwell time at certain Fermi
energy give rise to the peak structure of the admittance as
 clear in Eq.(\ref{gw-approx}). Meanwhile, the
role of driving electric field of high frequency $\Omega$ is to help
the incoming electron at the Fermi level jump into the resonant
levels in double dots via photo absorption or emission. Whenever the
Fermi level matches with the resonant energy plus integer multiple
of $\hbar\Omega$, the electron can dwell in the dots and the
admittance has peaks. This process is depicted in Fig. 3 a).

In Fig. 2 b), we show the noise power as a function of Fermi level.
The contribution from equilibrium Nyquist noise  is due to electron
states near Fermi level. At low frequency and zero temperature,
$S_{0}\approx
 \frac{e^2}{4\pi}\omega^3\tau_d^2(E_F)$\cite{buttiker} and $S_{P}(\omega)\approx \frac{e^{2}\hbar}{2\pi}\omega^{2}
 \sum_{m\neq 0}\int d\epsilon |\sum_{l} s^{*}_{l+m}(\epsilon)\frac{d}{d\epsilon}s_{l}(\epsilon+m\hbar\Omega)|^{2}
 f(\epsilon)(1-f(\epsilon+m\hbar\Omega))$.
The sharp peaks in Fig. 2(b) are attributed to the nonzero dwell
time at the Fermi levels. In contrast, we find that the
non-equilibrium part of noise $S_{P}(\omega)$ shows step-wise
behaviour. {\it Why does it show steps?}
 The electrons below Fermi level contribute to the
 non-equilibrium noise through photon absorption/emission.
Note that, in contrast to the case of the admittance, there is no
driving probe field of frequency $\omega$. Therefore, the noise
power is not necessarily a quantity for the Fermi level. The
electrons below Fermi level can contribute to the current
fluctuation via photo-assisted tunneling into the dots.

\begin{figure}
\includegraphics[width=0.6\figurewidth]{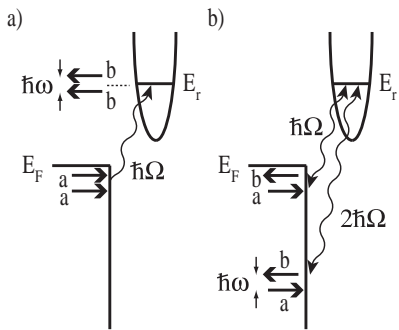}
 \caption{Schematic figures explaining the processes involved in
the admittance a) and the current noise b). See the text. }
\label{mechanism}
\end{figure}

The incoming electron states of energy $E<E_{F}$ contribute to the
noise when $E-E_{r}$ is the integer multiple of $\hbar\Omega$. (Fig.
3  b)). Since we consider the current noise power $S(\omega)$ at low
frequency $\omega<\Omega$, the outgoing electron states of the
energy other than $E+\hbar\omega$ are not involved in the low
frequency noise. Therefore, the number of pairs of incoming electron
of energy $E$ and the outgoing states of energy $E+\hbar\omega$
determine the strength of the current noise.(Fig. 3 b)). As the
Fermi level increases, the number of the pairs increases, which give
rise to the step structure. The step arises whenever the Fermi level
matches with the resonant energy plus integer multiple of
$\hbar\Omega$.

While there have been experimental
works on the electrical noise under ac excitation for diffusive
conductors\cite{schoelkopf} and tunnel
junctions\cite{gabelli2,deblock}, so far there has been no
experimental realization of the driven nano-structure tunnel-coupled
to single lead. To study the quantum aspect of the admittance
discussed in this work, the experimental system by Gabelli et.
al.\cite{gabelli} seems most relevant to the present theoretical
work where the dc conductance is zero. For experimental observation
of the resonant admittance peaks predicted in this work, the quantum
dot in use in Ref.\cite{gabelli} should be electrically driven and
smaller enough to ensure the dot's quantized energy spacing is
larger than the temperature energy scale. To detect the quantum
noise at high frequencies, the techniques in Ref.\cite{deblock,onac}
might be useful where the high frequency noise signal is converted
to dc current.

In conclusion, we investigate
 the low frequency admittance and current noise of nano-structure
which is driven by a high frequency field. A fluctuation-dissipation
relation for the driven system is obtained. The phase delay time
defined through Floquet scattering matrix is essential to understand
the admittance.
 The current noise power shows steps as a function of the Fermi energy
when the admittance shows peaks. The Fermionic nature of electrons
or the exchange correlation of the incoming electrons is important
to the step structure of the noise power.

\begin{acknowledgements}
 This work was supported by the Korea Science and Engineering
Foundation(KOSEF) grant funded by the Korea government(MOST)
(No.R01-2007-000-10837-0).
\end{acknowledgements}


\vspace{-0.5cm}
\begin{thebibliography}{99}
\bibitem{feve} G. F\'eve et. al.,Science {\bf 316}, 1169 (2007).
\bibitem{gabelli} J. Gabelli, et.al., Science {\bf 313}, 499 (2006).
\bibitem{blanter} Ya. M. Blanter and M. B\"uttiker, Phys. Rep. {\bf 336}, 1 (2000).
\bibitem{Levitov}G. B. Lesovik and L. S. Levitov Phys. Rev. Lett. {\bf 72},
538(1994).
\bibitem{Lamacraft} A. Lamacraft Phys. Rev. Lett. {\bf 91}, 036804 (2003).
\bibitem{camalet} S. Camalet, et.al., Phys. Rev. Lett. {\bf 90},
210602 (2003).
\bibitem{buttiker} M. B\"uttiker, A. Pr\^etre,
and H. Thomas, Phys. Rev. Lett. {\bf 70}, 4114 (1993); M.
B\"uttiker, H. Thomas, A. Pr\^etre, Phys. Letters A {\bf 180}, 364
(1993); A. Pr\^etre, and H. Thomas, and M. B\"uttiker, Phys. Rev. B,
{\bf 54}, 8130 (1996).
\bibitem{nigg} S. E. Nigg, R. L\'opez, and M.
B\"uttiker, Phys. Rev. Lett. {\bf 97}, 206804 (2006).
\bibitem{wigner} E. P. Wigner, Phys. Rev. {\bf 98}, 145 (1955);
Felix T. Smith, Phys. Rev. {\bf 118}, 1 (1960).
\bibitem{lesovik} G. B. Lesovik and R. Loosen, JETP Lett. {\bf 65}, 295 (1997).
\bibitem{ahn-physe} K.-H. Ahn, H. C. Park, B. Wu, Physica E. {\bf
34}, 468 (2006); K.-H. Ahn, J. Korean Phys. Soc., {\bf 47}, 666
(2005).
\bibitem{moskalets} M. Moskalets and M. B\"uttiker, Phys.
Rev. B, {\bf 66}, 205320 (2002).

\bibitem{schoelkopf} R. J. Schoelkopf, et. al., Phys. Rev. Lett. {\bf 80}, 2437
(1998).
\bibitem{gabelli2} J. Gabelli and B. Reulet, Phys. Rev. Lett. {\bf
100}, 026601 (2008).
\bibitem{deblock} R. Deblock, et. al., Science {\bf 301},
203 (2003).
\bibitem{onac} E. Onac, et. al., Phys. Rev. Lett. {\bf 96}, 176601 (2006).

\end{thebibliography}

\end{document}